\documentclass[fleqn,10pt]{wlscirep}
\usepackage[utf8]{inputenc}
\usepackage[T1]{fontenc}

\usepackage{graphicx} 

\usepackage{amsmath,amsthm,amssymb,mathrsfs}
\usepackage{bm}

\usepackage{color}

\title{Associative memory by virtual oscillator network based on single spin-torque oscillator}

\author[1]{Yusuke Imai}
\author[1*]{Tomohiro Taniguchi}
\affil[1]{National Institute of Advanced Industrial Science and Technology (AIST), Research Center for Emerging Computing Technologies, Tsukuba, Ibaraki 305-8568, Japan}

\affil[*]{tomohiro-taniguchi@aist.go.jp}

\begin{abstract}
A coupled oscillator network may be able to perform an energy-efficient associative memory operation. 
However, its realization has been difficult because inhomogeneities unavoidably arise among the oscillators during fabrication and lead to an unreliable operation. 
This issue could be resolved if the oscillator network were able to be formed from a single oscillator. 
Here, we performed numerical simulations and theoretical analyses on an associative memory operation that uses a virtual oscillator network based on a spin-torque oscillator. 
The virtual network combines the concept of coupled oscillators with that of feedforward neural networks. 
Numerical experiments demonstrate successful associations of $60$-pixel patterns with various memorized patterns. 
Moreover, the origin of the associative memory is shown to be forced synchronization driven by feedforward input, where phase differences among oscillators are fixed and correspond to the colors of the pixels in the pattern. 
\end{abstract}

\begin{document}

\flushbottom
\maketitle
%
%

The human brain has a sophisticated function called associative memory \cite{kohonen12}, whereby it can remember a pattern when shown a portion of that pattern. 
This function has been modeled in various ways with the goal of achieving a better understanding of brain activity and realizing energy-efficient bio-inspired computing. 
Since the development of an autocorrelation model in the 1970s \cite{nakano72,kohonen72,anderson72}, several theoretical models, such as the Hopfield model \cite{hopfield82}, have been developed that draw their inspiration from the characteristics of neural activity \cite{amari77,hemmen86,mceliece87,waugh90,morita93,yoshizawa93,bolle93,krebs99,mcgraw03,zhao04}. 
These models have also been implemented in experimental devices. 
For example, the associative memory operation was recently performed in a spintronic memory consisting of a nanometer-scale ferromagnetic multilayer \cite{borders17}. 
In addition to these efforts embodying neuronal dynamics, it has been proposed that synchronized phenomena in coupled oscillator networks can be used to perform the associative memory operation \cite{hoppensteadt97,hoppensteadt99,corinto07,mirchev13,maffezzoni15}. 
For example, a detailed analysis was conducted on an $LC$-circuit oscillator network performing the operation  \cite{maffezzoni15}. 
A network of spintronic oscillators, called spin-torque oscillators (STOs), has also been shown to perform an associative memory operation \cite{prasad22}.


There are two major issues with using an oscillator network for the associative memory operation. 
One is unstable operation due to inhomogeneity in the oscillator's parameters. 
For example, variations in frequency among the oscillators are unavoidable in experimental realizations; they prevent a synchronization between the oscillators and decrease the accuracy of the associative memory \cite{maffezzoni15}. 
The other issue is that the required number of oscillators grows with the amount of input data. 
There are numerous challenges in fabricating a large number of oscillators and getting them to interact with each other. 
These issues might be resolved if we can construct an oscillator network virtually by using a single physical oscillator \cite{tsunegi22}. 
Such a network would have no inhomogeneities in its parameters as only one oscillator would have to be fabricated. 
However, there are questions on how such a network could be realized and how it could show synchronization phenomena. 


In this work, we demonstrate an associative memory operation by a virtual oscillator network through numerical simulations and theoretical analyses. 
First, we provide a detailed description of the virtual oscillator network consisting of a single physical oscillator. 
In particular, we discuss the principles involved, i.e., those of the coupled oscillator networks and feedforward neural networks. 
Next, we show that a virtual oscillator network consisting of a single STO can recognize several different $60$-pixel patterns by numerically simulating the motion of the STO. 
We reveal that the feedforward input in the virtual network forces the virtual oscillators to synchronize and that this phenomenon results in the associative memory operation. 

\section*{Results}

\subsection*{Associative memory operation of this study}

\begin{figure}
\centerline{\includegraphics[width=0.8\columnwidth]{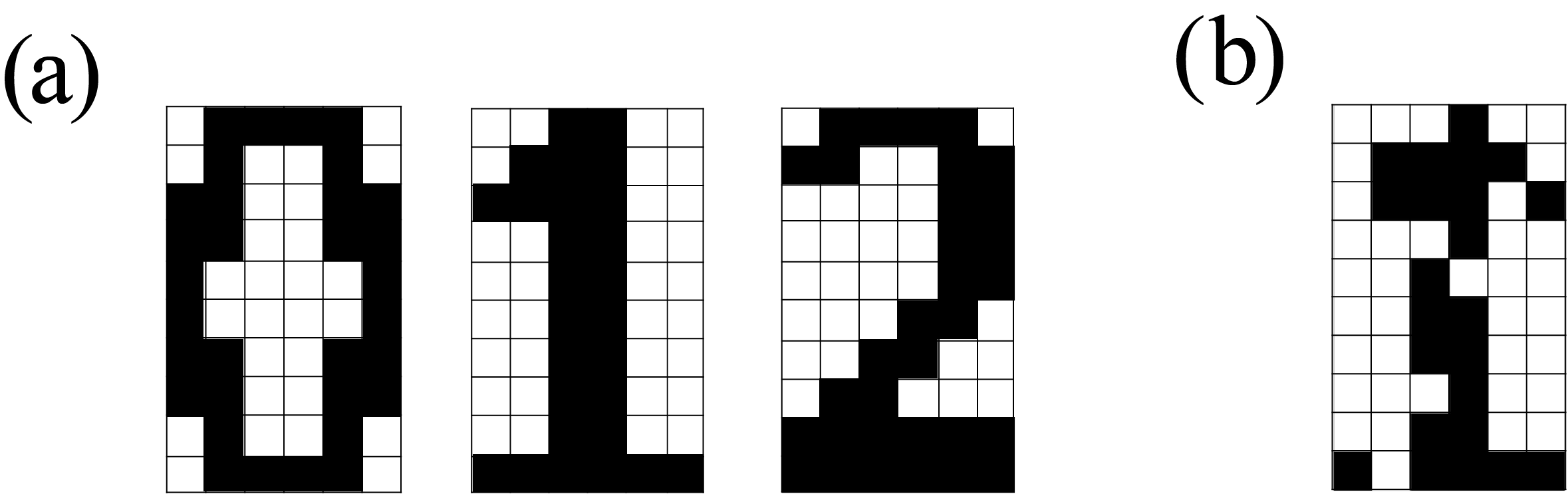}}
\caption{
            Examples of memorized patterns and a pattern to be recognized. 
            (a) Three ($N_{\rm m}=3$) memorized patterns, ``0'', ``1'', and ``2''. 
            (b) The pattern to be recognized resembles memorized pattern ``1''. 
            The oscillator network tries to associate the pattern to be recognized with the pattern ``1''. 
            In an associative memory operation performed by a system consisting of $N$ oscillators, the color of the $i$th ($i=1,2,\cdots,N$) pixel is determined by the phase $\psi_{i}$ of the corresponding oscillator. 
             The color is white (black) when the phase difference, $\Delta\psi_{i}=\psi_{i}-\psi_{1}$, is $0$  ($\pi$). 
             The color is on a gray scale when the phase difference is $0<\Delta\psi_{i}<\pi$. 
         \vspace{-3ex}}
\label{fig:fig1}
\end{figure}

The associative memory operation studied here is to associate a pattern, called the pattern to be recognized, with a pattern in a stored set of patterns, called memorized patterns.
For example, suppose that the three patterns, ``0'', ``1'', and ``2'', shown in Fig. \ref{fig:fig1}(a) are memorized, and the one shown in Fig. \ref{fig:fig1}(b) is the pattern to be recognized: we can see that the pattern to be recognized is similar to the memorized pattern ``1''. 
Throughout this paper, we will suppose the memorized patterns use $10$(rows)$\times$$6$(columns)$=60$-pixels patterns for memorized patterns and patterns to be recognized. 

In the following subsections, we describe the concept of our virtual oscillator network after briefly reviewing a conventional oscillator network for comparison. 
Then, we demonstrate through numerical simulations that the virtual oscillator network can perform the associative memory operation.  



\begin{figure}
\centerline{\includegraphics[width=1.0\columnwidth]{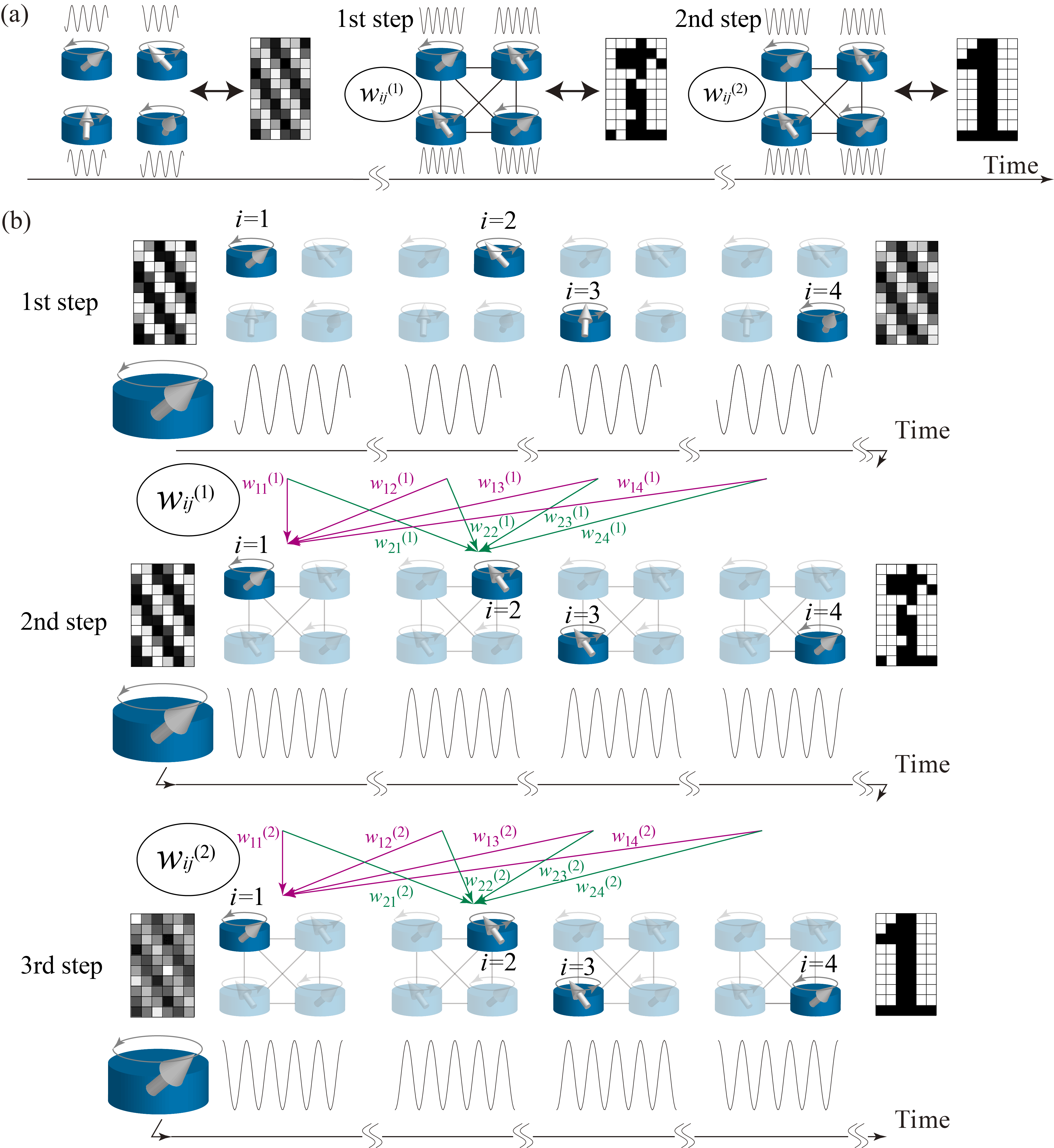}}
\caption{
             Schematic illustration of conventional and virtual oscillator networks. 
             (a) In the conventional oscillator network, the oscillators are initially uncoupled (left). 
             Therefore, the phase of each oscillator is arbitrary. 
             When the oscillators interact with appropriate weights [$w_{ij}^{(1)}$], the phases saturate to values corresponding to the pattern to be recognized (middle). 
             When the weight changes [$w_{ij}^{(2)}$], the phases change so that the corresponding pattern resembles one of memorized patterns (right). 
             (b) In a virtual oscillator network, we drive an oscillation of a single oscillator and divide its output into $N$ parts. 
             The $i$th part is regarded as an output from the $i$th virtual oscillator. 
             First [top of (b)], we measure the $N$ outputs. 
             The corresponding pattern in this step is arbitrary because there is no correlation among the oscillators. 
             Second [middle of (b)], an external force is added to the oscillator. 
             This force is a linear combination of the outputs in the first step with appropriated weights [$w_{ij}^{(1)}$]. 
             The phase of each part eventually saturates to a value corresponding to the pixel color in the pattern to be recognized. 
             Third [bottom of (b)], the second step is repeated while the force is a linear combination of the outputs in the second step with weights $w_{ij}^{(2)}$. 
             Eventually, the phases saturate to the values corresponding to the memorized pattern most resembling the pattern to be recognized. 
         \vspace{-3ex}}
\label{fig:fig2}
\end{figure}


%


\subsection*{Associative memory operation by conventional oscillator network}


The associative memory operation by a conventional coupled oscillator network consists of two steps \cite{maffezzoni15}.
The first step is to give a correspondence between the phases of the oscillators and the colors of the pattern to be recognized. 
We prepare $N$ oscillators corresponding to the pixels of the pattern to be recognized, where $N$ is the number of oscillators (pixels). 
We introduce phases $\psi_{i}$ ($i=1,\cdots,N$) and phase differences $\Delta\psi_{i}=\psi_{i}-\psi_{1}$. 
The color of the $i$th pixel is determined by $\cos\Delta\psi_{i}$, which is white (black) when $\Delta\psi_{i}=0$ ($\pi$). 
According to this definition, the color of the first pixel is always white (see also the Methods for the definitions of color). 
Initially, there are no interactions between the oscillators. 
Thus, their phases are arbitrary, and the colors in the pattern are random, as schematically shown on the left of Fig. \ref{fig:fig2}(a). 
When interactions between the oscillators are introduced and the interaction strengths are appropriately determined by the Hebbian rule, all the phase differences become $0$ or $\pi$ in correspondence with the white and black pixels of the pattern to be recognized, as shown in the middle of Fig. \ref{fig:fig2}(a) (see also Methods for model of the conventional oscillator network). 
Here, the Hebbian rule means that the interaction strength between the $j$th and $i$th oscillator is proportional to the weight,  
\begin{equation}
  w_{ij}^{(1)}
  =
  \xi_{i}^{\rm R}
  \xi_{j}^{\rm R},
  \label{eq:weight_1}
\end{equation} 
where $\xi_{i}^{\rm R}=+(-)1$ when the color of the pattern to be recognized at the $i$th pixel is white (black). 
Thus, $w_{ij}^{(1)}=+(-)1$ when the colors of the $i$th and $j$th are the same (opposite). 

The second step is to replace the weights by the following ones, which can be regarded as an average of the weights among the memorized patterns,  
\begin{equation}
  w_{ij}^{(2)}
  =
  \frac{1}{N_{\rm m}}
  \sum_{m=1}^{N_{\rm m}}
  \xi_{i}^{m}
  \xi_{j}^{m}, 
  \label{eq:weight_2}
\end{equation}
where $N_{\rm m}$ is the number of memorized patterns. 
The symbol $m=1,2,\cdots,N_{\rm m}$ is used to distinguish the memorized patterns. 
For example, the memorized patterns ``0'', ``1'', and ``2'' in Fig. \ref{fig:fig2}(a) are labelled $m=1$, $2$, and $3$. 
The parameter $\xi_{i}^{m}$ is $+(-)1$ when the color of the $i$th pixel in the $m$th memorized pattern is white (black). 
Then, the oscillator phases change to those of the memorized pattern most resembling the pattern to be recognized, and the association is achieved, as shown in the right in Fig. \ref{fig:fig2}(a). 


\subsection*{Description of associative memory operation by virtual oscillator network}


The associative memory operation by a virtual oscillator network consists of three steps. 

First, we measure an oscillation of a single oscillator and divide it into $N$ parts, as schematically shown on the first line of Fig. \ref{fig:fig2}(b). 
The $i$th part of the measured data is regarded as the output from the $i$th oscillator in a virtual network. 
In this step, the phase of each part is arbitrary, and therefore, the pattern arising from it is random. 
The measured data should be stored in a computer in order for it to be used in the next step. 

Second, we excite another oscillation and divide the measured data into $N$ parts again. 
At the initial time of each part, the phase, as well as the pattern determined from it, is arbitrary, as shown in the middle of Fig. \ref{fig:fig2}(b). 
This time, however, we apply an external force to the oscillator that is proportional to a linear combination of the measured data in the first step with weights (\ref{eq:weight_1}). 
For example, in this study, the external force comes from a torque excited by an external magnetic field, which applied during the $i$th part of the oscillation is given by 
\begin{equation}
  H_{i}^{(1)}  
  =
  \mathscr{H}
  \sum_{j=1}^{N}
  w_{ij}^{(1)}
  y_{j}^{(1)}, 
  \label{eq:interacting_field_1}
\end{equation}
where $\mathscr{H}$ denotes the amplitude and $y_{j}^{(1)}$ is the output from the $j$th oscillator measured in the first step [see also Methods for the detailed definition of $y_{j}^{(1)}$ in the numerical simulations]. 
Therefore, Eq. (\ref{eq:interacting_field_1}) is an oscillating function with the frequency of the oscillator. 
Because of the application of the magnetic field, the phase in each part eventually saturates to a certain value, and the pattern to be recognized is output, as shown in the middle of Fig. \ref{fig:fig2}(b). 
Note that the output signal of this process should be stored in a computer. 

Third, we perform a measurement similar to one in the second step but the magnetic field applied during the $i$th part is replaced by 
\begin{equation}
  H_{i}^{(2)}  
  =
  \mathscr{H}^{\prime}
  \sum_{j=1}^{N}
  w_{ij}^{(2)}
  y_{j}^{(2)}, 
  \label{eq:interacting_field_2}
\end{equation}
where $\mathscr{H}^{\prime}$ denotes the amplitude, while $y_{j}^{(2)}$ is the output from the $j$th oscillator measured at the second step (see also Methods pertaining to the numerical simulations). 
The weights $w_{ij}^{(2)}$ are given by Eq. (\ref{eq:weight_2}). 
The phase at the end of each part saturates to a value corresponding to the memorized pattern most resembling the pattern to be recognized, as shown in the bottom of Fig. \ref{fig:fig2}(b); i.e., the associative memory operation is completed. 

There are several differences between the conventional and virtual oscillator networks (see also Methods for the models). 
For example, the oscillators in the conventional oscillator network interact instantaneously, and their phase differences saturate to values corresponding to pixel colors as a result of mutual synchronization. 
On the other hand, the oscillators in the virtual oscillator network do not interact each other instantaneously. 
As can be seen in Eqs. (\ref{eq:interacting_field_1}) and (\ref{eq:interacting_field_2}), the oscillator outputs from the previous steps are used in the magnetic field in the current step. 
From perspective, the virtual oscillator network is similar to a feedforward neural network because the information on the oscillator phases in one step is sent to the oscillation in the next step. 
At the same time, we should note that the weights in the virtual oscillator network are fixed, as in the case of the conventional oscillator network. 
This is in contrast with a feedforward neural network used in deep learning, in which weights are updated by backpropagation. 
Thus, the virtual oscillator network can be regarded as a hybrid combination of a coupled oscillator network and a feedforward neural network. 
In the discussion below, we will reveal that the feedforward inputs cause forced synchronization among the divided parts and result in the associative memory operation. 
Before that, however, we must demonstrate that this virtual oscillator network can actually perform the associative memory operation.


\subsection*{Equation of motion of oscillator}

As an oscillator in the virtual oscillator network, we use a vortex STO, which has various advantages for practical applications and has been frequently used in spintronics experiments on bio-inspired computing \cite{locatelli14,grollier16,torrejon17,grollier20}. 
An STO consists of a ferromagnetic/nonmagnetic multilayer on the nanometer scale, as schematically shown in Fig. \ref{fig:fig3}(a). 
A vortex of magnetic moments appears when a diameter and thickness of a cylinder-shape ferromagnet are on the order of 100 and 1 nm, respectively. 
When an electric current and/or magnetic field are applied to the STO, magnetic moments show precessions around their equilibrium direction. 
According to a recent experiment on chaos excitation in an STO \cite{kamimaki21}, we assume that a force added to the virtual oscillator network corresponds to a torque excited by magnetic field, as mentioned above. 
It has been shown both experimentally and theoretically that the dynamics in a vortex STO are well described by the Thiele equation \cite{thiele73,guslienko06PRL,guslienko06,ivanov07,khvalkovskiy09,guslienko11,dussaux12,grimaldi14}, which is the equation of motion for a center of the vortex structure, called the vortex core (see also Methods for Thiele equation): 
\begin{equation}
\begin{split}
  &
  -G \mathbf{e}_{z}
  \times
  \dot{\mathbf{X}}
  -
  |D|
  \left(
    1
    +
    \xi 
    s^{2}
  \right)
  \dot{\mathbf{X}}
  -
  \kappa
  \left(
    1
    +
    \zeta 
    s^{2}
  \right)
  \mathbf{X}
  +
  a_{J} J p_{z}
  \mathbf{e}_{z}
  \times
  \mathbf{X}
  +
  c a_{J} J R_{0} 
  p_{x}
  \mathbf{e}_{x}
  +
  c \mu^{*}
  \mathbf{e}_{z}
  \times
  \mathbf{H}
  =
  \bm{0}, 
  \label{eq:Thiele}
\end{split}
\end{equation}
where $\mathbf{X}=(X,Y,0)$ represents the position of the vortex core in the $xy$ plane. 
While the physical meanings and the values of many parameters are explained in Methods, two quantities should be explained here. 
The first is the current density $J$, which causes a limit-cycle oscillation of the vortex core. 
The other is the external magnetic field $\mathbf{H}$, which is used to excite a torque. 
It is useful to notice that Eq. (\ref{eq:Thiele}) can be approximated as (see also Methods for the analytical solution of the Thiele equation)
\begin{equation}
  \dot{s}
  =
  a s
  -
  b s^{3}
  +
  \frac{c\mu^{*}}{GR}
  H_{y} 
  \sin\psi, 
  \label{eq:Thiele_radius}
\end{equation}
\begin{equation}
  \dot{\psi}
  =
  \frac{\kappa}{G}
  \left(
    1
    +
    \zeta 
    s^{2}
  \right)
  +
  \frac{c\mu^{*}}{GRs}
  H_{y}
  \cos\psi, 
  \label{eq:Thiele_phase}
\end{equation}
where $s=|\mathbf{X}|/R$ ($0 \le s \le 1$) is the distance of the vortex core from the center of the ferromagnet normalized by the disk radius $R$, while $\psi=\tan^{-1}(Y/X)$ is the phase. 
Here, $a=(|D|\kappa/G^{2})[(J/J_{\rm c})-1]$ and $b=(|D|\kappa/G^{2})(\xi+\zeta)$, where $J_{\rm c}=|D|\kappa/(Ga_{J}p_{z})$. 
The magnetic field $\mathbf{H}$ is assumed to have only a $y$ component $H_{y}$. 
Note that Eqs. (\ref{eq:Thiele_radius}) and (\ref{eq:Thiele_phase}) are similar to the equation of motion of the Stuart-Landau oscillator \cite{kuramoto03}. 
Therefore, the vortex core shows a limit-cycle oscillation around the disk center in the $xy$ plane with an oscillating amplitude $s_{0}=\sqrt{a/b}$ when $J$ exceeds a threshold value $J_{\rm c}$, while the terms related to $H_{y}$ act as a perturbation.  
The connection to such a fundamental nonlinear oscillator model indicates that our results are also valid for various oscillators in nature and engineering. 
Figure \ref{fig:fig3}(b) shows an example of nonperturbative vortex dynamics, showing an approximately circular oscillation of the vortex core around the disk center. 
The phase difference of the oscillation was used to define the colors in the patterns in the associative memory operation. 
Readers should note that the plots in Fig. \ref{fig:fig3}(b), as well as the results of the numerical simulations shown below, were obtained by solving Eq. (\ref{eq:Thiele}), while the approximate equations, Eqs. (\ref{eq:Thiele_radius}) and (\ref{eq:Thiele_phase}), are used in the model analyses described below. 


\subsection*{Demonstration of associative memory}


\begin{figure}
\centerline{\includegraphics[width=\columnwidth]{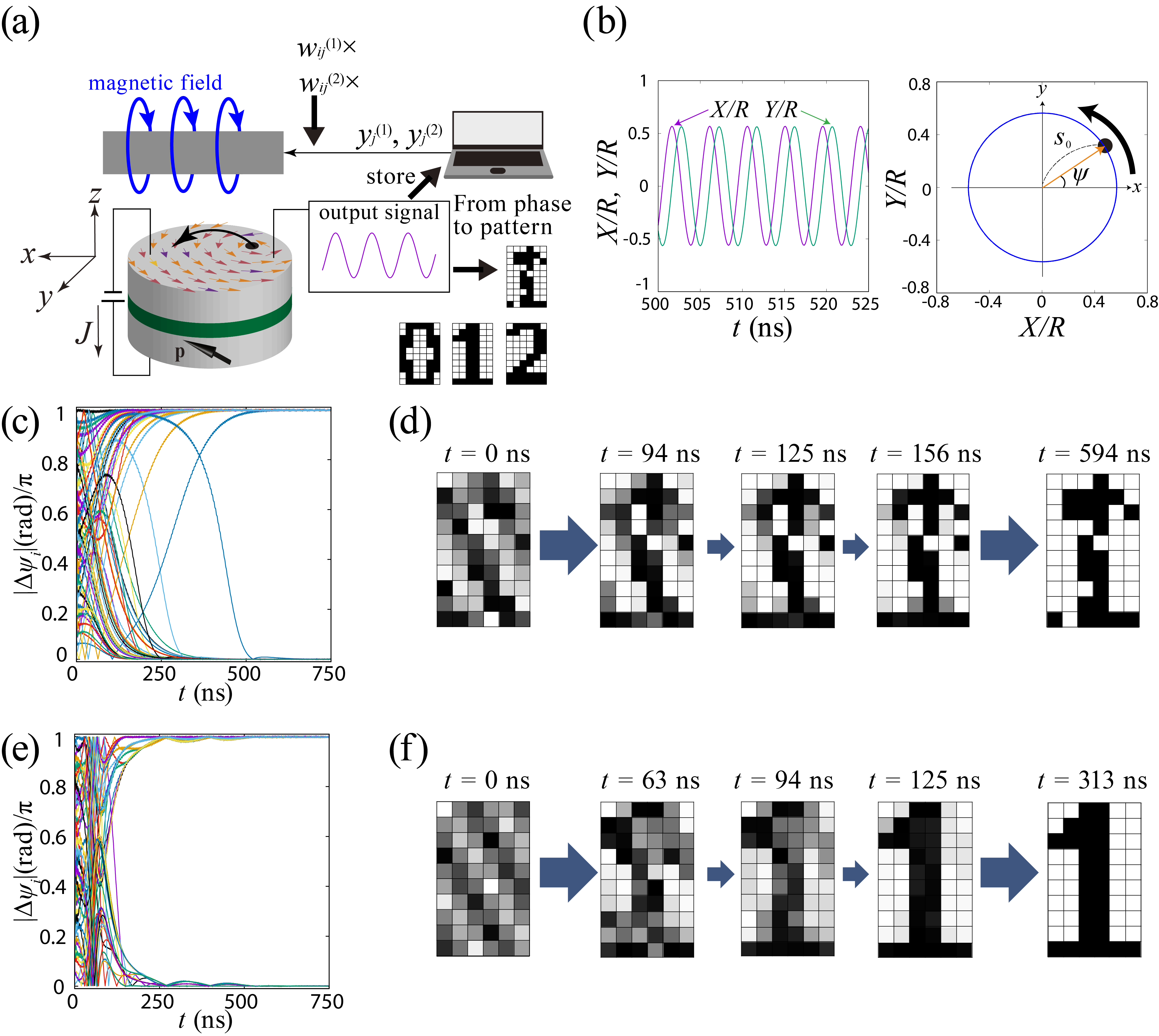}}
\caption{
             Description of STO and demonstration of associative memory by a virtual oscillator network.
             (a) Schematic illustration of vortex spin torque oscillator and (b) vortex-core dynamics driven by electric current. 
                  The STO has a cylindrical shape, and the $z$ axis is orthogonal to the circular plane. 
                  Magnetic moments, shown as colored arrows in top ferromagnet, form a circular structure. 
                  The black dot around which the moments turn is the vortex core. 
                  Electric current is injected into the STO; positive current flows from bottom to top in the figure.  
                  When the electric current density $J$ exceeds a threshold value, the vortex core oscillates around the disk center. 
                  The output signals from the STO during the first (second) step in Fig. \ref{fig:fig2}(b) are stored, and their linear combination with weights $w_{ij}^{(1)}$ [$w_{ij}^{(2)}$] defined from the pattern to be recognized (memorized patterns) is used as magnetic field during the second (third) step. 
                  For simplicity, the dynamics in the absence of the magnetic field is shown. 
                  The components of the vortex-core's position, $X/R$, and $Y/R$, oscillate around the disk center, and a trajectory is approximately a circle. 
                  The distance of the vortex-core's position from the disk center, $s$, is approximately constant value, $s_{0}$. 
                  The phase measured from the $x$ axis is denoted as $\psi$. 
             (c) Time evolutions of the $59$ phase differences, $\Delta\psi_{i}$ ($i=2,3,\cdots,60$) and (d) snapshots of generating a pattern to be recognized on $60$-pixels. 
             (e) Time evolutions of the phase difference and (f) snapshots of the corresponding pattern for association from memorized patterns.  
         \vspace{-3ex}}
\label{fig:fig3}
\end{figure}


\begin{figure}
\centerline{\includegraphics[width=0.75\columnwidth]{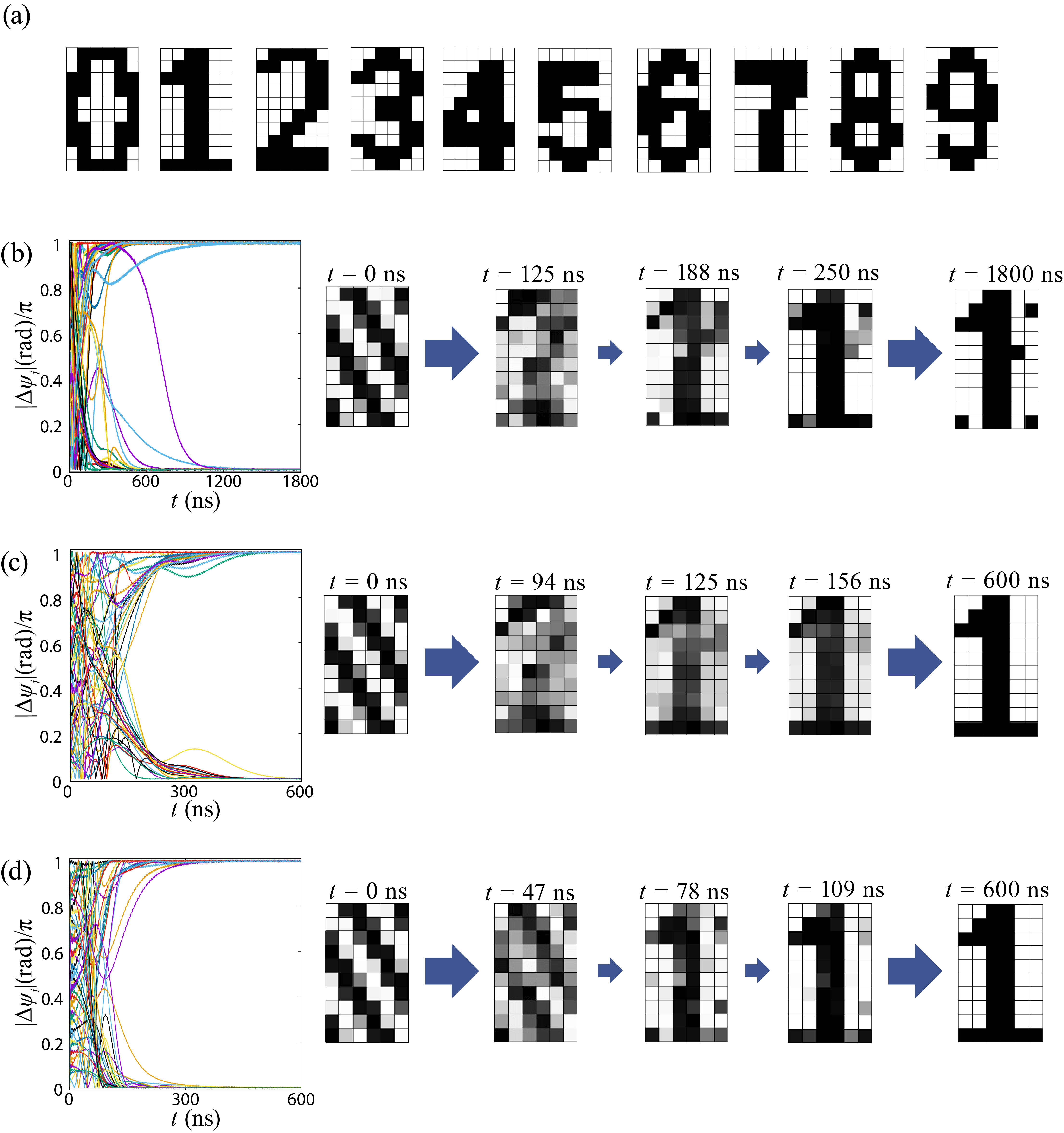}}
\caption{
             Problem of associative memory operation when the similarity between the memorized patterns is high and the number of patterns is large. 
             (a) Ten ($N_{\rm m}=10$) memorized patterns, ``0", ``1",$\cdots$,``9".
             (b) Time evolution of the phase difference during the association and snapshots of the corresponding pattern. 
                  In this case, the memorized patterns include both ``1'' and ``7''. 
                  Because of their similarity, the pattern does not finally saturate to ``1''. 
             (c) When ``7'' is removed from the memorized patterns ($N_{\rm m}=9$), the association is successful, even though there are nine remaining memorized patterns. 
             (d) The association is successful when the memorized patterns include only ``1'' and ``7''. 
         \vspace{-3ex}}
\label{fig:fig4}
\end{figure}


Figure \ref{fig:fig3}(c) shows the time evolution of the phase difference, $\Delta\psi_{i}$, obtained by solving Eq. (\ref{eq:Thiele}) with Eq. (\ref{eq:interacting_field_1}) substituting for $H_{y}$. 
Note that this solution corresponds to the second step in Fig. \ref{fig:fig2}(b). 
The phase differences saturate to $0$ or $\pi$ within a few hundred nanoseconds. 
Snapshots of patterns corresponding to this time evolution of the phases are shown in Fig. \ref{fig:fig3}(d). 
The patterns eventually settle to the one to be recognized. 
Figure \ref{fig:fig2}(b) shows the result of solving Eq. (\ref{eq:Thiele}) with Eq. (\ref{eq:interacting_field_2}) substituting for $H_{y}$. 
Here, Eq. (\ref{eq:weight_2}) in Eq. (\ref{eq:interacting_field_2}) is for the three memorized patterns in Fig. \ref{fig:fig1}(a). 
Figures \ref{fig:fig3}(e) and \ref{fig:fig3}(f) show the time evolution of the phase differences and snapshots of the corresponding patterns. 
Remind that the information of the phases corresponding to the colors of the pixels in the pattern to be recognized is included in the magnetic field in Eq. (\ref{eq:interacting_field_2}) through $y_{j}^{(2)}$. 
Consequently, even though the initial pattern is random, the oscillator phases finally saturate to values corresponding to one of the memorized patterns [Fig. \ref{fig:fig3}(f)]. 

The associative memory operation becomes more difficult when there are similar memorized patterns. 
To clarify this point, let us examine what happens when the number of the memorized patterns is increased, as shown in Fig. \ref{fig:fig4}(a) from the three in Fig. \ref{fig:fig1}(a). 
The added patterns do not affect the second step in Fig. \ref{fig:fig2}(b). 
For the association corresponding to the third step in Fig. \ref{fig:fig2}(b), the magnetic field, defined by Eq. (\ref{eq:interacting_field_2}), is changed by these new memorized patterns. 
As a result, the final pattern output resembles none of the memorized ones [Fig. \ref{fig:fig4}(b)]. 

This failure of the associative memory operation is due to two reasons. 
The first is that the pattern ``7'' is similar to the pattern ``1'', which should be the one associated. 
When ``7'' is excluded from the memorized patterns, the association succeeds, as shown in Fig. \ref{fig:fig4}(c). 
The second reason is that the number of memorized patterns is large. 
As shown in Fig. \ref{fig:fig4}(d), the association succeeds when the memorized patterns include only ``1'' and ``7'', the association is succeeded. 
Therefore, we conclude that an association may fail when the memorized patterns include similar patterns and the number of memorized patterns is large. 

To quantify the similarity between patterns $A$ and $B$, we introduce the degree of overlap: 
\begin{equation}
  \mathscr{O}(\bm{\xi}^{A},\bm{\xi}^{B})
  \equiv
  \frac{1}{N}
  \bigg|
    \sum_{i=1}^{N}
    \xi_{i}^{A}
    \xi_{i}^{B}
  \bigg|, 
  \label{eq:overlap}
\end{equation}
where $\bm{\xi}^{A}=(\xi_{1}^{A},\cdots,\xi_{N}^{A})$ is defined from the color of the $i$th pixel of  pattern $A$ [$\xi_{i}^{A}=+(-)1$ when the $i$th pixel is white (black)]. 
The overlap becomes $1$ when the two patterns are completely identical or their black and white colors are all exchanged (see also Methods for the definitions of color and overlap). 
For example, in the example shown in Figs. \ref{fig:fig1} and \ref{fig:fig3}, the degree of overlap between the pattern to be recognized and the memorized pattern ``0'' is $\mathscr{O}(\bm{\xi}^{\rm R},\bm{\xi}^{1})=18/60= 0.30$. 
It is$\mathscr{O}(\bm{\xi}^{\rm R},\bm{\xi}^{2})=44/60\simeq 0.73$ for pattern ``1'', and $\mathscr{O}(\bm{\xi}^{\rm R},\bm{\xi}^{3})=6/60=0.10$ for pattern ``2'' (the memorized patterns are labelled as $m=1,2,3,\cdots$ while the examples of memorized patterns in this work are ``0'', ``1'', ``2'', etc; thus, the label $m$ and the corresponding number are off by one). 
Since the degree of overlap of the pattern to be recognized and ``1'' is large in the examples in Figs. \ref{fig:fig1} and \ref{fig:fig3}, pattern ``1'' should be associated in this case. 
On the other hand, in the example shown in Fig. \ref{fig:fig4}, the overlap between the pattern to be recognized [Fig. \ref{fig:fig1}(b)] and ``7'' is also relatively large, i.e., $\mathscr{O}(\bm{\xi}^{\rm R},\bm{\xi}^{8})=32/60\simeq0.53$. 
In addition, the overlap between the memorized patterns ``1'' and ``7'', $\mathscr{O}(\bm{\xi}^{2},\bm{\xi}^{8})=28/60\simeq 0.47$, is also relatively large compared with those between the other patterns; for example, the overlap between ``1'' and ``8'' is $\mathscr{O}(\bm{\xi}^{2},\bm{\xi}^{9})=2/60\simeq 0.03$ (see also Supplementary Information, where the overlaps of the ten memorized patterns are summarized). 
Accordingly, when the memorized patterns include ``1'' and ``7'', the virtual oscillator network cannot associate a correct pattern, and the final pattern produced corresponds to none of the memorized ones. 
Similarly, when the number of memorized patterns is large, there might be patterns having large overlaps and the association fails. 

In summary, we have shown that the virtual oscillator network based on the algorithm in Fig. \ref{fig:fig2}(b) can perform the associative memory operation. 
Its accuracy, however, is low when the memorized patterns include some patterns having large overlaps and there is a large number of memorized patterns. 
Note that the maximum number of patterns that can be memorized by neural network is approximately $N/(2 \log N)$ \cite{mceliece87}. 
It would be of interest if such a formula can be derived for virtual oscillator networks in future. 

We examined the associative memory operation for various cases, i.e., for different patterns to be recognized, and studied the rate of the accurate association; see Supplementary Information.


\section*{Discussion}


Here we discuss the principles of the associative memory operation analytically by using Eqs. (\ref{eq:Thiele_radius}) and (\ref{eq:Thiele_phase}). 
As mentioned above, the operation consists of three steps, and in each step, the oscillator output is divided into $N$ parts. 
In what follows, we denote the phase of the vortex core during the $i$th part of the $k$th step as $\psi_{i}^{(k)}$. 
We also assume that the oscillation amplitude $s_{0}$ is approximately constant because the current density is fixed. 
Therefore, the oscillation frequency, $f=\varOmega/(2\pi)=[\kappa/(2\pi G)](1+\zeta s_{0}^{2})$, is also approximately constant (see also Methods for the analytical solution of the Thiele equation). 

The phase in the second step obeys, 
\begin{equation}
  \dot{\psi}_{i}^{(2)}
  =
  \varOmega
  +
  \frac{c\mu^{*}}{GRs_{0}}
  \mathscr{H}
  \sum_{\ell=1}^{N}
  \xi_{i}^{\rm R}
  \xi_{\ell}^{\rm R}
  y_{\ell}^{(1)}
  \cos\psi_{i}^{(2)}. 
  \label{eq:Thiele_phase_second_step}
\end{equation}
Thus, the phase difference between the $i$th and $j$th parts obeys, 
\begin{equation}
  \dot{\psi}_{i}^{(2)}
  -
  \dot{\psi}_{j}^{(2)}
  =
  \frac{c\mu^{*}}{GRs_{0}}
  \mathscr{H}
  \left(
    \sum_{\ell=1}^{N}
    \xi_{\ell}^{\rm R}
    y_{\ell}^{(1)}
  \right)
  \left(
    \xi_{i}^{\rm R}
    \cos\psi_{i}^{(2)}
    -
    \xi_{j}^{\rm R}
    \cos\psi_{j}^{(2)}
  \right). 
  \label{eq:equation_phase_difference}
\end{equation}
The steady state condition on the phase difference leads to 
\begin{equation}
  \xi_{i}^{\rm R}
  \cos\psi_{i}^{(2)}
  -
  \xi_{j}^{\rm R}
  \cos\psi_{j}^{(2)}
  =
  0. 
  \label{eq:condition_initilization}
\end{equation}
Note that $\xi_{i}^{\rm R}=+(-)1$ when the color at the $i$th pixel of the pattern to be recognized is white (black). 
Therefore, $\psi_{i}^{(1)}$ and $\psi_{j}^{(1)}$ will be in-phase $\psi_{i}^{(1)}=\psi_{j}^{(1)}$ [anti-phase $\psi_{i}^{(1)}=\psi_{j}^{(1)}\pm\pi$] when the colors of the $i$th and $j$th pixels are the same (opposite). 
As a result, the phase differences in the second step saturate to $0$ or $\pi$ corresponding to the white or black in the pattern to be recognized. 
Note that this synchronization is caused by a feedforward input from the first step, which corresponds to the second term on the right-hand side in Eq. (\ref{eq:Thiele_phase_second_step}). 
Here, the term $\sum_{\ell=1}^{N}\xi_{\ell}^{\rm R}y_{\ell}^{(1)}$ in Eq. (\ref{eq:Thiele_phase_second_step}) is the sum of the $N$ oscillator outputs $y_{\ell}^{(1)}$ in the first step, multiplied by the factor $\xi_{\ell}^{\rm R}$ determining the pixel color of the pattern to be recognized, and is common for all $i$ of Eq. (\ref{eq:Thiele_phase_second_step}). 
Equation (\ref{eq:Thiele_phase_second_step}) also includes a factor $\xi_{i}^{\rm R}$, which determines the sign of the input. 
Regarding these facts, the feedforward input has only two values, depending on the value of $\xi_{i}^{\rm R}$.
The phase synchronization among the $N$ parts in the second step is the result of forced synchronization with respect to this feedforward input, and the phase difference has only two values, $0$ or $\pi$, depending on the value of $\xi_{i}^{\rm R}$. 
This mechanism is in contrast with that of the previous work \cite{maffezzoni15}, where a mutual synchronization is the origin of the associative memory operation. 
Also, the method is different from the previous works \cite{romera18,skowronski19}. 
In Ref. \cite{romera18}, a forced synchronization of frequency with respect to an external signal was studied, while the input signal in the present work is generated by the oscillator output itself and the phase synchronization plays the central role in the associative memory operation. 
In Ref. \cite{skowronski19}, a delayed-feedback was used to generate input signal, while the input signal in the present work is generated by multiplying appropriated weight to perform the associative memory operation. 


We also note that, when $y_{\ell}^{(1)}$ is a simple trigonometric function, its linear combination, $\sum_{\ell=1}^{N}\xi_{\ell}^{\rm R}y_{\ell}^{(1)}$, is also a trigonometric function with the same frequency and a different phase. 
According to the above discussion, the phase of the term $\sum_{\ell=1}^{N}\xi_{\ell}^{\rm R}y_{\ell}^{(1)}$ does not play any role to excite forced synchronization among the $N$ parts. 
Thus, the term $\sum_{\ell=1}^{N}\xi_{\ell}^{\rm R}y_{\ell}^{(1)}$ could be replaced by, for example, $y_{1}^{(1)}$. 
In this case, it is unnecessary to measure other $(N-1)$ $y_{\ell}^{(1)}$ ($\ell=2,3,\cdots,N$) in the first step in Fig. \ref{fig:fig2}(b), although we solved the equation of motion for $N$ virtual oscillators to clarify similarities and differences between the second and third step. 
When $(N-1)$ parts in the first step are omitted for simplicity, the power consumption to drive the oscillator in the virtual oscillator network is proportional to $2N+1$, where $2N$ comes from the second and third steps in Fig. \ref{fig:fig2}(b). 
On the other hand, the power consumption in the conventional oscillator network is proportional to $2N$ because $N$ oscillators are driven two times, as implied in Fig. \ref{fig:fig2}(a). 
For a large $N$, the power consumption of two oscillator networks are comparable. 
The time required for the operation increases linearly as $N$ increases, which is not suitable for practical applications, although the same might be true for a conventional oscillator network because the relaxation time of the phase will depend on the number of the oscillators. 
the  time of a conventional (coupled) oscillator network might also increase as $N$ increases. 
However, the virtual oscillator network has an advantage from a viewpoint of reliability, as discussed below.


Next, we focus on the third step, where the phase during the $i$th part obeys
\begin{equation}
  \dot{\psi}_{i}^{(3)}
  =
  \varOmega
  +
  \frac{c\mu^{*}}{GRs_{0}}
  \mathscr{H}^{\prime}
  \frac{1}{N_{\rm m}}
  \sum_{m=1}^{N_{\rm m}}
  \sum_{\ell=1}^{N}
  \xi_{i}^{m}
  \xi_{\ell}^{m}
  y_{\ell}^{(2)}
  \cos\psi_{i}^{(3)}. 
  \label{eq:Thiele_phase_third_step_1}
\end{equation}
Since the oscillators in the second step are in the synchronized state, the output $y_{\ell}^{(2)}$ can be expressed as $y_{\ell}^{(2)}=\xi_{\ell}^{\rm R}\xi_{1}^{\rm R}y_{1}^{(2)}$, where $y_{1}^{(2)}$ is the output of the first part in the second step. 
We substitute this relation into Eq. (\ref{eq:Thiele_phase_third_step_1}) and assume that 
\begin{equation}
  \sum_{\ell=1}^{N}
  \xi_{\ell}^{m}
  \xi_{\ell}^{\rm R}
  \simeq
  \delta_{m,\mathscr{A}}
  \sum_{\ell=1}^{N}
  \xi_{\ell}^{m}
  \xi_{\ell}^{\rm R}, 
  \label{eq:assumption_delta}
\end{equation}
where the symbol $\mathscr{A}$ corresponds to a pattern in the memorized patterns that resembles the pattern to be recognized. 
The assumption (\ref{eq:assumption_delta}) means that only a pattern having a large degree of overlap with the pattern to be recognized contributes to the feedforward input. 
The other memorized patterns, which are greatly different from the pattern to be recognized, do not contribute to the feedforward input because of their small overlap. 
When the assumption is satisfied, Eq. (\ref{eq:Thiele_phase_third_step_1}) becomes 
\begin{equation}
  \dot{\psi}_{i}^{(3)}
  =
  \varOmega
  +
  \frac{c\mu^{*}}{GRs_{0}}
  \mathscr{H}^{\prime}
  \frac{1}{N_{\rm m}}
  y_{1}^{(2)}
  \xi_{1}^{\rm R}
  \left(
    \sum_{\ell=1}^{N}
    \xi_{\ell}^{\mathscr{A}}
    \xi_{\ell}^{\rm R}
  \right)
  \xi_{i}^{\mathscr{A}}
  \cos\psi_{i}^{(3)}. 
  \label{eq:Thiele_phase_third_step_2}
\end{equation}
Equation (\ref{eq:Thiele_phase_third_step_2}) is similar to Eq. (\ref{eq:Thiele_phase_second_step}), and therefore, the steady-state condition of the phase difference between the $i$th and $j$th parts in the third step is given by  
\begin{equation}
  \xi_{i}^{\mathscr{A}}
  \cos\psi_{i}^{(3)}
  -
  \xi_{j}^{\mathscr{A}}
  \cos\psi_{j}^{(3)}
  =
  0. 
  \label{eq:condition_association}
\end{equation}
Equation (\ref{eq:condition_association}) means that in-phase or anti-phase synchronization between the $N$ parts occurs, and the phase differences in the third step saturate to $0$ or $\pi$ corresponding to the white or black colors in a memorized pattern most resembling the one to be recognized. 


The operation principle is based on Eq. (\ref{eq:assumption_delta}). 
Equation (\ref{eq:assumption_delta}) is satisfied if there is only one pattern that has a large degree of overlap with the pattern to be recognized.  
On the other hand, if there are other patterns having large overlaps with the pattern to be recognized, Eq. (\ref{eq:assumption_delta}) is not satisfied. 
In this case, Eq. (\ref{eq:condition_association}) is not necessarily satisfied, and the colors in the steady state in the third step might be different from the pattern most resembling the one to be recognized or they might be gray (neither black nor white); see also Supplementary Information. 


Our analysis also assumed that the oscillation frequencies of the $N$ parts are the same. 
This assumption is a natural one because each part is obtained from a single oscillator. 
Technically speaking, the oscillation frequency in each part is varied by changing the magnitude of the electric current. 
If the oscillation frequencies of the $i$th and $j$th parts, denoted as $\varOmega_{i}/(2\pi)$ and $\varOmega_{j}/(2\pi)$, are different, the right-hand side of Eq. (\ref{eq:equation_phase_difference}) has an additional term $\varOmega_{i}-\varOmega_{j}$. 
In such a case, the phase difference is not well defined because $\psi_{i}$ and $\psi_{j}$ oscillate with different frequencies. 
Even if we introduce an instantaneous phase by, for example, making a Hilbert transformation, as was done in experiments \cite{tsunegi19}, the phase difference still does not necessarily saturate to $0$ or $\pi$. 
In such a case, the associative memory operation fails.  
Therefore, there is no reason to change the oscillation frequency in each part. 
This fact also indicates an advantage to using the virtual oscillator network. 
In the conventional oscillator network, variations in the oscillation frequency naturally appear because inhomogeneities in the parameters of the oscillators are unavoidable, and such variations lead to the failure of the associative memory operation \cite{maffezzoni15}. 
The virtual oscillator network does not have such variation and thus would be a more reliable associative memory. 
A weak point of the present proposal is, on the other hand, that the method requires a computer to store the output signal in each step, which is not preferable for practical applications. 
We would like to keep this issue as a future work. 




In conclusion, we described the concept of the associative memory operation by a virtual oscillator network and performed numerical simulations. 
The operation consists of three steps, where the output of one step is sent to the next step with weights defined by the Hebbian rule. 
In this sense, the virtual oscillator network can be regarded as a hybrid combination of a coupled oscillator network and a feedforward neural network. 
The network successfully associated black-and-white patterns with a few memorized patterns. 
However, it failed to make an association when the number of memorized patterns was large (ten compared to three) and some of the memorized patterns resembled each other. 
We also developed a theoretical analysis and clarified that the origin of the associative memory operation is forced synchronization driven by feedforward input. 
Either in-phase or anti-phase synchronization was excited among the oscillators and provides appropriate correspondence between the oscillator phases and the colors in the patterns. 
The virtual oscillator network is more reliable than a conventional oscillator network, which is affected by unavoidable inhomogeneities among the oscillators. 




\section*{Methods}


\subsection*{Definitions of color and overlap}

By convention, the first pixel (the pixel in the top left-hand corner of a pattern) is always white. 
The pattern should be regarded as the same even when all of the black and white pixels are swapped for each other, 
Mathematically, this means that $\sum_{i=1}^{N}\xi_{i}^{A}\xi_{i}^{B}=N$ when the patterns $A$ and $B$ are completely the same, and $\sum_{i=1}^{N}\xi_{i}^{A}\xi_{i}^{B}=-N$ when patterns $A$ and $B$ represent the same pattern but their black and white colors are completely swapped. 
According to this definition of the same figure, the maximum number of the difference between two patterns is $N/2$; in this case, the degree of overlap is zero (see also the discussion on noise in Supplementary Information).


\subsection*{Models of conventional and virtual oscillator networks}

The conventional oscillator network for the associative memory operation \cite{maffezzoni15} is based on the Kuramoto model \cite{kuramoto03}. 
The Kuramoto model describes the oscillator dynamics with a generalized phase, $\theta$. 
Moreover, the oscillators interact instantaneously, and the phase of the $i$th oscillator obeys
\begin{equation}
  \dot{\theta}_{i}
  =
  \omega
  +
  \mathscr{Q}
  \sum_{j=1}^{N}
  w_{ij}
  \sin
  \left(
    \theta_{i}
    -
    \theta_{j}
  \right),
  \label{eq:Kuramoto_model}
\end{equation}
where $\omega/(2\pi)$ is the oscillation frequency while $\mathscr{Q}$ is the interaction strength. 
For simplicity, we will assume that all oscillators share the same values of $\omega$ and $\mathscr{Q}$. 
The weight $w_{ij}$ is given by Eq. (\ref{eq:weight_1}) or (\ref{eq:weight_2}) depending on the step of the procedure.
In the $LC$-circuit model \cite{maffezzoni15}, $\mathscr{Q}w_{ij}$ is proportional to the transconductance. 
The phase difference between the $i$th and $j$th oscillators obeys 
\begin{equation}
  \dot{\theta}_{i}
  -
  \dot{\theta}_{j}
  =
  \mathscr{Q} 
  \left[
    \sum_{\ell=1}^{N}
    w_{i\ell}
    \sin\left(\theta_{i}-\theta_{\ell}\right)
    -
    \sum_{\ell=1}^{N}
    w_{j\ell}
    \sin\left(\theta_{j}-\theta_{\ell}\right)
  \right].
  \label{eq:equation_phase_difference_Kuramoto}
\end{equation}
In a limiting case of only two oscillators ($N=2$), the phase difference obeys 
\begin{equation}
  \dot{\theta}_{1}
  -
  \dot{\theta}_{2}
  =
  2 \mathscr{Q} 
  w_{12}
  \sin\left(\theta_{1}-\theta_{2}\right), 
\end{equation}
and the in-phase (anti-phase) synchronization of $\theta_{1}$ and $\theta_{2}$ is a stable fixed point when $\mathscr{Q}w_{12}$ is negative (positive).
The phase differences of $\theta_{i}-\theta_{j}=0,\pi$ are always fixed points even when there are a large number of oscillators $(N \ge 3)$. 
Accordingly, the phase differences in the conventional oscillator network saturate to the in-phase or anti-phase state, which thereby enables the associative memory operation. 

In the presence of frequency variations, the right-hand side of Eq. (\ref{eq:equation_phase_difference_Kuramoto}) has an additional term $\omega_{i}-\omega_{j}$. 
In this case, the phase difference is not stabilized, and this instability leads to an inaccurate associative memory operation \cite{maffezzoni15}.  

The Thiele equation is slightly different from the Kuramoto model in the following ways. 
First, the Thiele equation uses the phase $\psi$, which describes the vortex core's position in the $xy$ plane, instead of a generalized phase. 
This is because the quantity measured in experiments is the vortex core's position, and the phase synchronization studied in the experiments \cite{tsunegi19} corresponds to that of $\psi$, not a generalized phase $\theta$. 
Note that we can introduce a generalized phase analytically as $\theta=\psi+[\zeta\kappa/(Gb)]\ln(s/s_{0})$ with a phase sensitivity function $\mathbf{Z}=(-\sin\theta+[\zeta\kappa/(Gb)]\cos\theta,\cos\theta+[\zeta\kappa/(Gb)]\sin\theta,0)/s_{0}$. 
The analysis is mostly unchanged with the generalized phase, so we decided to use $\psi$ for simplicity.
Second, the equation of motion for the phase difference, Eq. (\ref{eq:equation_phase_difference}), includes a term $\cos\psi_{i}-\cos\psi_{j}$ whereas the Kuramoto model often uses an interacting term proportional to $\sin(\theta_{i}-\theta_{j})$. 
More generally, the interaction term in the Kuramoto model can be assumed to be a function of the phase difference, $\theta_{i}-\theta_{j}$ after applying an averaging technique with respect to a fast variable (see Ref. \cite{kuramoto03} for details). 
The difference between the two models might however be insignificant; notice that, by using formulas, $\cos x-\cos y=-2 \sin[(x+y)/2] \sin[(x-y)/2]$ and $\cos x+\cos y = 2\cos[(x+y)/2]\cos[(x-y)/2]$ and applying the averaging technique, the interaction term in our model can be approximated as a function of $\theta_{i}-\theta_{j}$. 
Third, as mentioned above, the input term in the virtual oscillator network consists of the oscillator output from the previous step, while the interaction in the Kuramoto model is instantaneous. 
Because of these differences, the associative memory operation by the virtual oscillator network is significantly different from those of conventional coupled oscillator networks on which previous experiments and the theoretical analyses have been conducted. 


\subsection*{Parameters in the Thiele equation}

Spin torque oscillators (STOs) mainly consist of a ferromagnetic metal/insulating layer/ferromagnetic metal trilayer. 
The first ferromagnetic layer of the trilayer is called the free layer and is where the magnetic vortex forms. 
The second ferromagnetic layer having a uniform magnetization is called the reference layer. 
When electric current is injected into STOs, spin-transfer torque \cite{slonczewski96,berger96,slonczewski05} is excited on the magnetic moments in the free layer and drives their dynamics \cite{dussaux12,grimaldi14}. 
The output signal from the STOs depends on the relative angle between the magnetizations in the free and reference layers. 

The definitions and physical meanings of the parameters in Eq. (\ref{eq:Thiele}) are as follows. 
The parameters $G=2\pi pML/\gamma$ and $D=-(2\pi\alpha ML/\gamma)[1-(1/2)\ln(R_{0}/R)]$ consist of the polarity $p(=\pm 1)$ of the vortex core, the saturation magnetization $M$, the thickness $L$ of the ferromagnet, the gyromagnetic ratio $\gamma$, the Gilbert damping constant $\alpha$, and the vortex radius $R_{0}$. 
The chirality $c(\pm 1)$ of the vortex core also appears in Eq. (\ref{eq:Thiele}). 
The parameters $\kappa$ and $\zeta$ relate to a magnetic potential energy defined as $W=(\kappa/2)[1+(\zeta/2)s^{2}]|\mathbf{X}|^{2}$. 
The dimensionless parameter $\xi$ is introduced to describe the nonlinear damping in a highly excited state \cite{dussaux12}. 
The parameter $\kappa$ relates to the material parameters as $\kappa=(10/9)4\pi M^{2}L^{2}/R$ \cite{dussaux12}. 
The parameter $a_{J}=\pi\hslash P/(2e)$ includes the reduced Planck constant $\hslash$, spin polarization $P$ of the electric current, and the elementary charge $e(>0)$. 
The vector $\mathbf{p}=(p_{x},0,p_{z})$ is the unit vector pointing in the magnetization direction in the reference layer. 
Here, we assume that $\mathbf{p}$ lies in the $xz$ plane, by convention. 
As a result, the output signal from the vortex STO is proportional to the $y$ component of the vortex core's position. 
The parameter $\mu^{*}$ is $\pi MLR$. 

The material parameters used in this study were taken from typical experiments and simulations \cite{dussaux12,grimaldi14,tsunegi21}: $M=1300$ emu/cm${}^{3}$, $\gamma=1.764\times 10^{7}$ rad/(Oe s), $\alpha=0.01$, $L=5$ nm, $R=187.5$ nm, $R_{0}=10$ nm, $P=0.7$, $\xi=2.0$, and $\zeta=0.1$. 
The polarity and chirality were assumed to be $p=+1$ and $c=+1$, for simplicity. 
The magnetization direction in the reference layer was $\mathbf{p}=(\sin 60^{\circ},0,\cos 60^{\circ})$. 
An electric current $I$ of $1$ mA corresponded to a current density $J$ of $0.9$ MA/cm${}^{2}$. 
The electric current in the numerical simulations was set to 4.0 mA. 

We do not include field-like torque in the Thiele equation, which is expressed as $-cb_{J}JRp_{x}\mathbf{e}_{y}$ in Eq. (\ref{eq:Thiele}); see, for example, Ref. \cite{imai22}. 
This is because its magnitude was not visible in an experiment using CoFeB/MgO based STO \cite{tsunegi22}. 
One might consider to inject the input through the field-like torque, instead of the torque due to the external magnetic field as we have done. 
However, the modulation of the field-like torque requires that of electric current, which leads to the modulation of the frequency of the STO. 
Since the advantage of our proposal is that the frequency is unique during the operation, we do not prefer to use the field-like torque for injecting the input.


\subsection*{Analytical solution of the Thiele equation}

The Gilbert damping constant $\alpha$ is often small, in such cases, $|D|/G \simeq \alpha \ll 1$. 
Also, the radius $R_{0}$ of the vortex core is much shorter than the disk radius, $R$. 
Therefore, by neglecting terms related to $R_{0}$ and higher-order terms of $\alpha$, we can approximate Eq. (\ref{eq:Thiele}) as Eqs. (\ref{eq:Thiele_radius}) and (\ref{eq:Thiele_phase}) in terms of $s=|\mathbf{X}|/R$ and $\psi=\tan^{-1}(Y/X)$. 
The approximated Thiele equation without magnetic field is 
\begin{equation}
  \dot{s}
  =
  a s 
  -
  b s^{3},
\end{equation}
\begin{equation}
  \dot{\psi}
  =
  \frac{\kappa}{G}
  \left(
    1
    +
    \zeta s^{2}
  \right). 
\end{equation}
These equations are identical to the Stuart-Landau equation \cite{kuramoto03}, which was introduced by Landau to describe the evolution of turbulence phenomenologically and was derived from hydrodynamics by Stuart. 
This equation provides one of the simplest example of Hopf bifurcation. 
A stable solution of $s$ is $s_{0}=\sqrt{a/b}$ ($0$) for $a>(<)0$, or equivalently, $J/J_{\rm c}>(<)1$. 
When $J/J_{\rm c}>1$, i.e., the current density $J$ exceeds a threshold value $J_{\rm c}$, the vortex core oscillates around the disk center with an oscillation amplitude $s_{0}$ and the frequency $f=[\kappa/(2\pi G)](1+\zeta s_{0}^{2})$. 
Note that the oscillation frequency is proportional to the current density $J$ through the term $s_{0}^{2}=a/b$ ($a\propto J$), which has been confirmed by both experiments and simulations \cite{dussaux12,grimaldi14}. 
Even in the presence of the magnetic field, the oscillation frequency remains $f$, if the input strength is weak. 

The solution of $s$ obtained from the exact Thiele equation, Eq. (\ref{eq:Thiele}), shows a small oscillation around $s_{0}$ \cite{yamaguchi20}. 
This means that the trajectory of a limit-cycle oscillation is approximately circular but also has a small amplitude modulation. 
This small modulation is caused by the term $ca_{J}JR_{0}p_{x}\mathbf{e}_{x}$ in Eq. (\ref{eq:Thiele}), which breaks the axial symmetry of the dynamics around the $z$-axis. 
The deviation of $s$ from $s_{0}$ is, however, negligible, and the oscillation trajectory is approximately circular, as shown in Fig. \ref{fig:fig3}(b). 
Therefore, it is reasonable to omit the term from Eqs. (\ref{eq:Thiele_radius}) and (\ref{eq:Thiele_phase}). 
Note that this term arises from the in-plane component $p_{x}$ of the magnetization in the reference layer. 
$p_{x}$ plays a role in experiments for the following reason. 
Recall that the output signal measured in experiments depends on the relative angle of the magnetizations in the free and reference layers. 
Since the vortex core is located in the $xy$ plane, a finite $p_{x}$ is necessary to detect its position. 
On the other hand, the $z$ component $p_{z}$ is also necessary because the spin-transfer torque originating from it excites the limit-cycle oscillation of the vortex core. 
In fact, the threshold current density $J_{\rm c}=|D|\kappa/(Ga_{J}p_{z})$ is inversely proportional to $p_{z}$; therefore, if $p_{z}$ is zero, $J_{\rm c}$ becomes infinite and the oscillation cannot be excited. 
In experiments \cite{tsunegi19,kamimaki21}, the magnetization initially pointed in an in-plane direction, where $p_{z}=0$. 
A finite $p_{z}$ was induced by applying an external magnetic field in the $z$ direction. 

According to Eqs. (\ref{eq:Thiele_radius}) and (\ref{eq:Thiele_phase}), one might consider that the magnetic field changes the value of $s$ from $s_{0}$ and modifies the oscillation frequency. 
Such a frequency shift is, however, negligibly small, which can be discussed accordingly. 
First, remind that the frequency of the magnetic field applied during the second step is the frequency of the vortex core without the magnetic field because it consists of the output during the first step. 
The fact that the phases in the second step are saturated to $0$ or $\pi$, as shown in Fig. \ref{fig:fig3}(c), indicates that the forced phase synchronization occurs, and the frequency of the vortex core in the second step is the same with that in the first step. 
Second, let us roughly estimate the frequency shift by the application of the magnetic field. 
The change of $s$ by the magnetic field will be maximized when the phase of the magnetic field $H_{y}$ in Eq. (\ref{eq:Thiele_radius}) is the same with $\psi$. 
In this case, the magnitude of the last term in Eq. (\ref{eq:Thiele_radius}), averaged over a precession period $\tau=1/f$, is about $[c\mu^{*}/(2GR)]H_{y}\tau\sim (\gamma/2)\mathscr{H}\tau$. 
The period $\tau$ is about $5$ ns while $\mathscr{H}$ is on the order of $1$ Oe; see next section. 
Accordingly, the shift $\Delta s$ of $s$ by the application of the magnetic field is less than $0.1$ at maximum. 
As mentioned, the oscillation frequency is proportional to $1+\zeta s^{2}$. 
Using $\zeta=0.1$ and $s_{0}\simeq 0.6$, estimated from Fig. \ref{fig:fig3}(b), the frequencies with and without $\Delta s$, which are proportional to $1+\zeta s_{0}^{2}$ and $1+\zeta (s_{0}+\Delta s)^{2}$, respectively, differ only $1$ \% at maximum. 
Therefore, we consider that the frequency modulation by the application of the magnetic field is negligible. 

One might be of interested in the applicability of the Thiele equation. 
While the original Thiele equation assumes a translation symmetry in an infinite space, a finite-size effect of nanostructured may restrict the applicability of the equation. 
Therefore, the Thiele equation had been applied to analyses on small-amplitude dynamics \cite{goto11}. 
There have been, at the same time, several efforts to make the equation applicable to large-amplitude dynamics. 
For example, adding nonlinear frequency and damping terms is one approach \cite{dussaux12,grimaldi14}, which is also used in the present work, where the additional terms are characterized by the dimensionless parameters $\xi$ and $\zeta$. 
Adding further higher-order nonlinear terms is also investigated recently \cite{araujo22,wergifosse22arXiv,araujo22arXiv}. 
It was also shown that the Thiele equation is applicable to analyze small-amplitude dynamics, and effort has been made to extrapolating it to a large-amplitude dynamics, such as vortex-core expulsion, although there are some limitations \cite{jenkins16}.
In the present study, we use the model developed in Refs. \cite{dussaux12,grimaldi14} due to the following reasons. 
First, the applicability of the model to wide ranges of parameters has been verified by comparison with experiments \cite{dussaux12,grimaldi14,tsunegi21}.  
Second, adding higher-order nonlinear terms does not change main conclusion in this work. 
These terms might change, for example, current dependence of the oscillation frequency. 
In the present work, however, the frequency is kept constant, and thus, adding such terms do not play a central role in the associative memory operation. 
Third, the Thiele equation with the present approximation clarifies the connection between spintronics and other research fields such as nonlinear science and computer science. 
This is because the equation can be reduced to the Stuart-Landau equation, as mentioned above. 
The Stuart-Landau equation has a long history, as in the case of the Thiele equation, and has been frequently used in nonlinear science \cite{kuramoto03,pikovsky03}. 
The present work indicates that the Stuart-Landau oscillator can be emulated in nanostructures and therefore, prompts communications between spintronics and other research fields. 
Therefore, although we understand that there have been great efforts \cite{araujo22,wergifosse22arXiv,araujo22arXiv,guslienko14} for the validity and applicability of the Thiele equation, we use the model developed in Refs. \cite{dussaux12,grimaldi14}. 
Note that the Oersted field generated in the current, discussed in these previous works, does not play a role in the associative memory operation because the current magnitude is kept constant during the operation. 
Also, since the external magnetic field induces forced synchronization, a frequency shift due to an external magnetic field studied in the previous work \cite{araujo22} does not exist in the present algorithm. 



\subsection*{Details of the numerical simulations}

The associative memory operation in the virtual oscillator network consists of three steps. 
The initial state of the vortex core in each step is prepared by adding a thermal activation to the Thiele equation and solving it in the absence of magnetic field, as is done in Ref. \cite{imai22}. 
The torque due to the thermal activation gives an additional term, $-\eta_{x}\mathbf{e}_{x}-\eta\mathbf{e}_{y}$, to the left-hand side of Eq. (\ref{eq:Thiele}), which obeys the fluctuation-dissipation theorem, 
\begin{equation}
  \langle \eta_{i}(t) \eta_{j}(t^{\prime}) \rangle 
  =
  2k_{\rm B}T |D| \delta_{ij}
  \delta(t-t^{\prime}), 
\end{equation}
where the temperature $T$ is $300$ K. 
The solution of the Thiele equation in each step is divided into $N=60$ parts, where the time width of each part is denoted as $\tilde{t}$. 
In the experiment \cite{tsunegi22}, a certain time period was inserted between these parts to remove their correlation. 
In contrast, our numerical simulations used parallel computations, wherein the initial state of each part was randomly prepared using the method described above. 
The value of $\tilde{t}$ was changed depending on the number of memorized patterns, as well as the number of noisy pixels in the pattern to be recognized.
For example, $\tilde{t}$ is $750$ ns in Fig. \ref{fig:fig3}(c). 
For all cases, $\tilde{t}$ was divided into $\tilde{n}=\tilde{t}/t_{\rm p}$ parts, where $t_{\rm p}=0.125$ ns. 

Now let us explain the meanings of $y_{\ell}^{(1)}$ and $y_{\ell}^{(2)}$ in Eqs. (\ref{eq:interacting_field_1}) and (\ref{eq:interacting_field_2}). 
Since they are defined in a similar manner, we will describe only $y_{\ell}^{(1)}$. 
When defining the magnetic field in Eq. (\ref{eq:interacting_field_1}), it is convenient to reset the time origin for each part; i.e., each of the $N$ parts runs from $t=0$ to $t=\tilde{t}$. 
Remember that the output from the STO is proportional to the $y$ component of the vortex core's position, $Y$. 
We denote the solution of the normalized $y$ component, $y=Y/R$ ($0 \le y \le 1$), during the $\ell$th part in the first step as $y_{\ell}$. 
Then, $y_{\ell}^{(1)}$ is made from $y_{\ell}$ as follows, 
\begin{equation}
  y_{\ell}^{(1)}
  =
  \sum_{n=0}^{\tilde{n}-1}  
  y_{\ell}(nt_{\rm p})
  \left\{
    \Theta(t-n t_{\rm p})
    -
    \Theta[t-(n+1)t_{\rm p}]
  \right\}, 
  \label{eq:input_y_def}
\end{equation}
where $\Theta(t)$ is a step function. 
Note that $\Theta(t-nt_{\rm p})-\Theta[t-(n+1)t_{\rm p}]$ is $1$ for $nt_{\rm p} \le t < (n+1)t_{\rm p}$ and is zero for the other times; thus, it has a pulse shape. 
Equation (\ref{eq:input_y_def}) means that the input strength is constant for $nt_{\rm p} \le t < (n+1)t_{\rm p}$ and is proportional to $y_{\ell}(t)$ at $t=nt_{\rm p}$. 
$\tilde{n}$ is the number of input pulses. 
There are two reasons to shape the output $y$ into a pulse. 
The first one relates to numerical simulations. 
In this work, the Thiele equation was solved within a time increment of $\Delta t=0.005$ ns, which is shorter than the pulse width $t_{\rm p}$. 
It was, however, impractical to store the output at each $\Delta t$ step because the amount would have been huge. 
Second, there is a technical limitation in real experiments on a measurable time step. 
The value we used, $t_{\rm p}=0.125$ ns, is close to shortest possible time step in an experiment \cite{tsunegi22}. 
Because of these reasons, we define $y_{\ell}^{(1)}$ used in the magnetic field, Eq. (\ref{eq:interacting_field_1}), as a pulse input. 
At the same time, we emphasize that $t_{\rm p}$ is much shorter than an oscillation period of the vortex core, $1/f=4.48$ ns ($f=223$ MHz). 
In addition, the pulse-shaped $y_{\ell}^{(1)}$s are continuously injected. 
Therefore, the magnetic field can be approximately regarded as a continuously oscillating signal with respect to the STO. 

The strength of the input $\mathscr{H}$ in the second step is $1.0$ Oe, while that in the third step is $\mathscr{H}^{\prime}=N_{\rm m}\times 0.2$ Oe. 
Here, we increase $\mathscr{H}^{\prime}$ as the number $N_{\rm m}$ of memorized patterns increases. 
This is because the time necessary to reach a steady state becomes long as $N_{\rm m}$ increases; therefore, to perform the numerical simulations efficiently, the input strength should be made to increase with $N_{\rm m}$.







\section*{Acknowledgements}

The authors are grateful to Sumito Tsunegi and Yuki Hibino for discussion. 
Some of the results were obtained from a project 
``Innovative AI Chips and Next-Generation Computing Technology Development/(2) Development of Next-generation Computing Technologies/Exploration of Neuromorphic Dynamics towards Future Symbiotic Society'' commissioned by NEDO. 
T. T. was supported by a JPS KAKENHI Grant, Number 20H05655.


\section*{Author contributions statement}

T.T. designed the project. 
Y.I. performed the numerical simulations based on a program code developed by T.T. 
Y.I. also developed a model analysis with help of T.T. 
Y.I. and T.T. wrote the manuscript and prepared the figures.


\section*{Competing interests}

The authors declare no competing interests. 


\section*{Data availability}

The datasets used and/or analyses during the current study available from the corresponding author on reasonable request.


\section*{Additional information}

\textbf{Supplementary information} is available for this paper. 
\\
\textbf{Correspondence} and requests for materials should be addressed to T.T.





\end{document}


\flushbottom
\maketitle



%
%


\section*{Degree of overlap between the numerical patterns}

Table S\ref{tab:tab1} lists the overlap degrees between the ten memorized patterns, ``0'', ``1'', $\cdots$, ``9'', used in the main text. 
Remind that we discuss the similarly between the patterns ``1'' and ``7'' in the main text. 
As can be seen in the table, the pattern having the largest overlap with respect to ``1'' is certainly ``7''. 

There are other pairs of numbers having large degrees of overlap.  
For example, the overlap between ``3'' and ``9'' is 0.54, which is larger than that between ``1'' and ``7''. 
Moreover ``8'' has large overlaps with several other patterns; 
the overlap with ``3'', ``5'', ``6'', and ``9'' are 0.73, 0.50, 0.73, and 0.60, respectively. 
Recall that we concluded in the main text that the associative memory operation fails when the number of the memorized patterns is large and some of the memorized patterns have large overlaps. 
We also showed that the association becomes accurate when the number of the memorized patterns is reduced or similar patterns are removed from the memorized ones. 
In the next section, we show that these conclusions are valid even for the memorized pattern ``8''. 


\begin{table}
\caption{Table of overlaps between ten memorized patterns, ``0'', ``1'', $\cdots$, ``9''.}
\label{tab:tab1}
\centering
\scalebox{1.3}[1.3]{
\begin{tabular}{|c|cccccccccc|}  \cline{1-11}
overlap &           0 &           1 &           2 &           3 &           4 &           5 &           6 &           7 &           8 &           9 \\ \cline{1-11}
0 &        1.00 &        0.37 &        0.07 &        0.27 &        0.13 &        0.17 &        0.27 &        0.30 &        0.20 &        0.27 \\
1 &        0.37 &        1.00 &        0.03 &        0.10 &        0.03 &        0.13 &        0.10 &        0.47 &        0.03 &        0.10 \\
2 &        0.07 &        0.03 &        1.00 &        0.13 &        0.20 &        0.30 &        0.27 &        0.17 &        0.07 &        0.13 \\
3 &        0.27 &        0.10 &        0.13 &        1.00 &        0.13 &        0.43 &        0.47 &        0.30 &        0.73 &        0.53 \\
4 &        0.13 &        0.03 &        0.20 &        0.13 &        1.00 &        0.23 &        0.20 &        0.10 &        0.33 &        0.07 \\
5 &        0.17 &        0.13 &        0.30 &        0.43 &        0.23 &        1.00 &        0.70 &        0.07 &        0.50 &        0.43 \\
6 &        0.27 &        0.10 &        0.27 &        0.47 &        0.20 &        0.70 &        1.00 &        0.10 &        0.73 &        0.60 \\
7 &        0.30 &        0.47 &        0.17 &        0.30 &        0.10 &        0.07 &        0.10 &        1.00 &        0.03 &        0.10 \\
8 &        0.20 &        0.03 &        0.07 &        0.73 &        0.33 &        0.50 &        0.73 &        0.03 &        1.00 &        0.60 \\
9 &        0.27 &        0.10 &        0.13 &        0.53 &        0.07 &        0.43 &        0.60 &        0.10 &        0.60 &        1.00 \\ \cline{1-11}
\end{tabular}
}
\end{table}


\section*{Associative memory operation for the pattern of ``8"}

\begin{figure}
\centerline{\includegraphics[width=0.6\columnwidth]{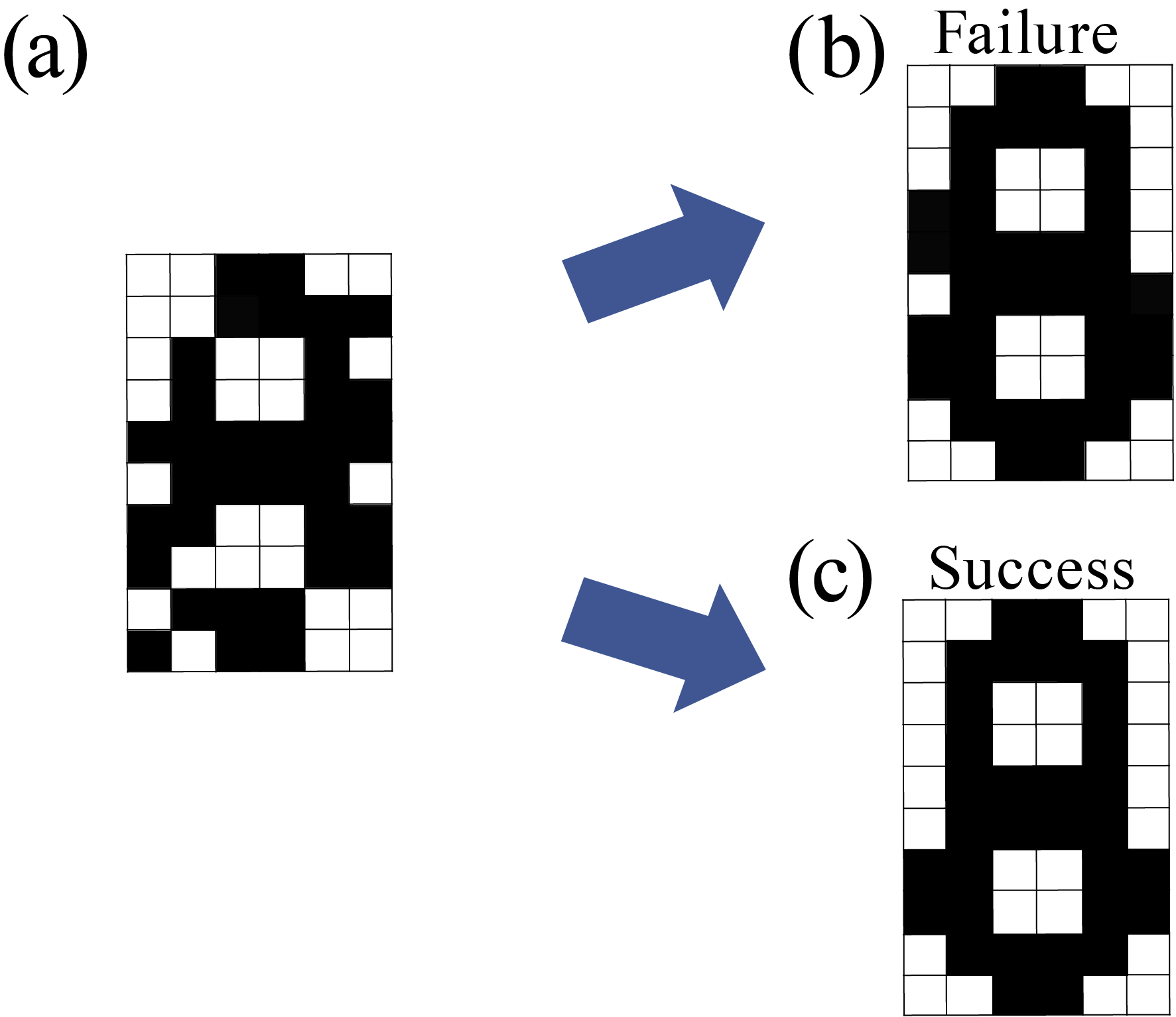}}
\caption{
            (a) The pattern to be recognized was obtained by adding noise to the pattern ``8''.  
            (b) When the memorized patterns included ten patterns, the association of ``8'' failed. 
            (c) When the memorized patterns included only ``3'' (or ``6'') and ``8'', the association succeeded, even though these patterns have a large overlap. 
         \vspace{-3ex}}
\label{fig:figS1}
\end{figure}

Here, we made a pattern to be recognized by adding noise to the pattern ``8'' as shown in Fig. S\ref{fig:figS1}(a), and then examined the associative memory operation with it. 
The amplitude of the magnetic field during the association was $\mathscr{H}^{\prime}=N_{\rm m}\times 0.17$ Oe. 
First, we used ten memorized patterns, ``0'', ``1'', $\cdots$, ``9''. 
The finally obtained pattern is shown in Fig. S\ref{fig:figS1}(b); it is slightly different from ``8'' and does not match any of the memorized patterns exactly. 
This result is similar to the one shown in the main text, where the pattern to be recognized was similar to ``1'' but had noise in it, and the finally obtained pattern did not correspond to any of the memorized patterns when the number of the memorized patterns was ten. 
Next, we examined the association by removing some of the memorized patterns. 
For example, when we used only ``3'' and ``8'' as the memorized patterns, the finally obtained pattern eventually becomes ``8'', as shown in Fig. S\ref{fig:figS1}(c); thus, the association succeeded. 
Moreover, the association succeeded even when we used ``6'' instead of ``3''. 
These results support the conclusions in the main text. 


\begin{figure}
\centerline{\includegraphics[width=1.0\columnwidth]{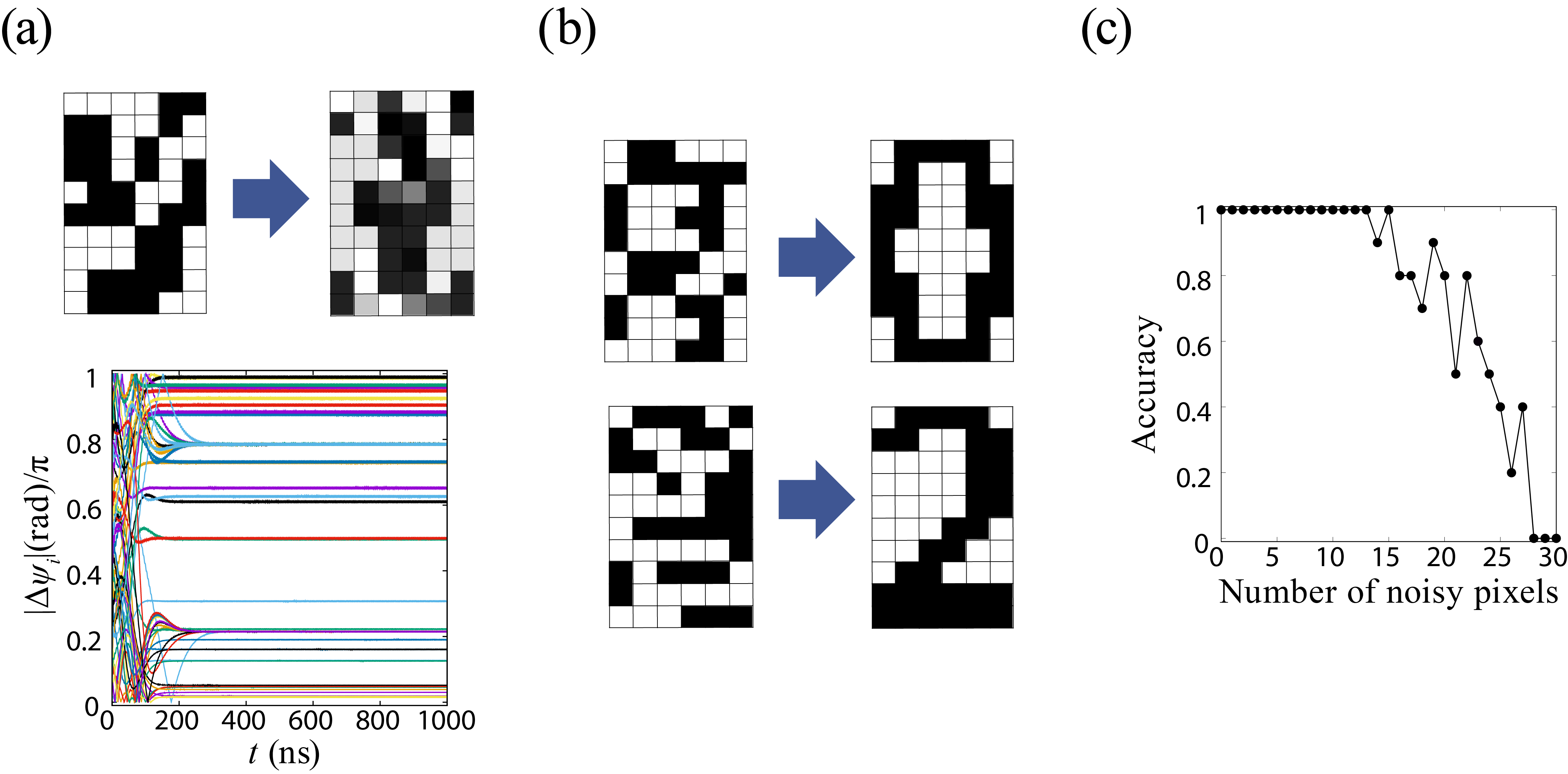}}
\caption{
            (a) Example of an inaccurate association. 
                The top left and right patterns are the pattern to be recognized and the final pattern obtained after the association. 
                In this case, the final pattern does not match any of the memorized ones (``0'', ``1'', and ``2''). 
                In particular, some of its pixels are colored neither black nor white. 
                This means that some phase differences did not saturate to $\pi$ or $0$, as shown in the bottom figure. 
            (b) Other examples of inaccurate associations. 
                 In these cases, the finally associated patterns are ``0'' and ``2'', even though the patterns to be recognized were generated by adding noise to ``1''. 
            (c) Dependence of accuracy of the associative memory operation on the number of noisy pixels.
         \vspace{-3ex}}
\label{fig:figS2}
\end{figure}


\section*{Dependence of associative memory operation on noise}

Here, we investigate the dependence of the accuracy of the associative memory operation on the number of noisy pixels in the pattern. 
Before showing the results, however, we should define what we mean by noise (or noisy pixel) and accuracy. 

Noise is defined as swapping black and white colors in a memorized pattern to make a new pattern to be recognized. 
For example, the pattern to be recognized in Fig. S\ref{fig:figS1}(a) is obtained by swapping the colors of $8$ pixels from the pattern in Fig. S\ref{fig:figS1}(c); therefore, there are $8$ noisy pixels in this case. 
If the virtual oscillator network can associate the pattern in Fig. S\ref{fig:figS1}(a) with that Fig. S\ref{fig:figS1}(c), we say it is an accurate association. 
If the finally obtained pattern still includes some noise, as in the case of Fig. S\ref{fig:figS1}(b), it is classified as an inaccurate association. 

Here, one might imagine other definitions of accuracy. 
As mentioned above, we prepared the patterns to be recognized by adding a noise to one of the memorized patterns. 
Let us call the pattern to be recognized $R$ and the original memorized pattern $A$. 
If the amount of noisy pixel is large, $R$ might become more similar to one of the other memorized patterns, call it pattern $B$. 
Mathematically, this means that the overlap between $R$ and $B$ is larger than that between $R$ and $A$, even though the pattern $R$ is generated from the pattern $A$. 
In this case, the finally associated pattern might be $B$. 
Similarly, even when the overlap between $R$ and $A$ is the larger than those between $R$ and any of the other memorized patterns, the finally obtained pattern might become one of the other memorized patterns. 
We regarded  such associations to be inaccurate, even though the finally obtained pattern is amongst the memorized ones. 
This is because noise reduction (or pattern recovery) remains a challenging aspect of the associative memory operation, so we aimed to associate $R$ with the pattern it is derived from, i.e., $A$. 

With these definitions of noise (noisy pixel) and accuracy in hand, let us evaluate the accuracy as a function of the number of noisy pixels. 
Here, let us consider associating ``1'' from three memorized patterns, ``0'', ``1'', and ``2'', as in the case studied in the main text. 
As mentioned in the Methods, the maximum number of noisy pixels is $N/2$, where $N$ is the number of the pixels, $60$ in the present study.  

Figure S\ref{fig:figS2}(a) shows an example of the association in which the pattern to be recognized was generated by adding $30$ noisy pixels to pattern ``1''. 
In this case, the finally associated pattern is not ``1''. 
Furthermore, some of the pixels are neither black nor white. 
This means that the phase differences of some parts did not saturate to either $\pi$ or $0$ [see Fig. S\ref{fig:figS2}(a)]. 
Such a gray pattern is obtained when the number of noisy pixels is relatively large; in contrast, when the number of noisy pixels is relatively small, the finally obtained pattern is black-and-white even if the association fails, as shown in Fig. S\ref{fig:figS1}(b). 

Figure S\ref{fig:figS2}(b) shows other examples for which the patterns to be recognized were obtained by adding $30$ noisy pixels to ``1''. 
The finally associated patterns are other memorized patterns, ``0'' and ``2''. 
Here, the overlaps between the patterns to be recognized and ``0'' and ``2'' are relatively small; the overlaps between the patterns to be recognized and the finally obtained patterns were $0.1$. 
Even for such a small overlap, the association became inaccurate because the overlaps between the patterns to be recognized and ``1'' are zero. 
Moreover, the finally obtained pattern has the largest overlap with the pattern to be recognized. 
For example, in the top example in Fig. S\ref{fig:figS2}(b), the overlaps between the pattern to be recognized and the memorized patterns are $0.1$ for ``0'', $0$ for ``1'', and $0.03$ for ``2''. 
Similarly, in the bottom example in Fig. S\ref{fig:figS2}(b), the overlaps are $0.03$ for ``0'', $0$ for ``1'', and $0.1$ for ``2''. 
Therefore, the finally associated pattern is the one with the largest overlap. 
According to the definition of accuracy above, however, these examples should be regarded as inaccurate associations. 

Figure S\ref{fig:figS2}(c) summarizes the accuracy as a function of the number of noisy pixels. 
Here, we quantified the accuracy as follows. 
We prepared ten ($N_{\rm e}=10$) patterns to be recognized for a certain number of noisy pixels, where noisy pixel was added randomly to the ten patterns. 
For example, the patterns shown in Figs. S\ref{fig:figS2}(a) and S\ref{fig:figS2}(b) are examples of such patterns with $30$ noisy pixels. 
The accuracy in Fig. S\ref{fig:figS2}(c) is the number of the correctly associated patterns divided by $N_{\rm e}=10$. 
Therefore, the accuracy is $1$ when the association succeeds for $N_{\rm e}=10$ prepared patterns to be recognized, while the accuracy is zero when the association fails for all prepared patterns. 
We used ``0'', ``1'', and ``2'' as the memorized patterns and generated the patterns to be recognized by adding noisy pixels to pattern ``1''. 
The accuracy was $1$ for relatively small number of noisy pixels. 
It deviated from $1$ when the number of noisy pixels was close to half of the maximum value, i.e., the association was inaccurate when the number of noisy pixels was approximately larger than $N/4$. 
The accuracy was zero when the number of noisy pixels was close to the maximum number, $N/2$, because the overlap between the pattern to be recognized and the original pattern becomes approximately zero.







